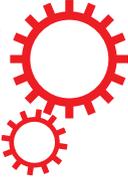

# Microwave control of the superconducting proximity effect and minigap in magnetic and normal metals

Jacob Linder, Morten Amundsen & Jabir Ali Ouassou

We demonstrate theoretically that microwave radiation applied to superconducting proximity structures controls the minigap and other spectral features in the density of states of normal and magnetic metals, respectively. Considering both a bilayer and Josephson junction geometry, we show that microwaves with frequency $\omega$ qualitatively alters the spectral properties of the system: inducing a series of resonances, controlling the minigap size $E_{mg}$, and even replacing the minigap with a strong peak of quasiparticle accumulation at zero energy when $\omega = E_{mg}$. The interaction between light and Cooper pairs may thus open a route to active control of quantum coherent phenomena in superconducting proximity structures.

Combining materials with different properties is a certain way to generate exciting physics at their interface. Superconducting hybrid structures are particularly interesting in this regard due to the coherent quantum correlations that give rise to dissipationless transport of both charge and, when combined with magnetic materials, spin. There is currently much interest in discovering ways to exert well-defined control the properties of such proximity structures, including the electronic density of states, the critical temperature at which superconductivity arises, and the appearance of supercurrents[1–3].

The influence of microwave radiation on superconductors has been studied in several works, and includes investigations of its effect on the critical superconducting current[4], the dissipative conductivity[5], the current-phase relation in Josephson junctions[6,7], the non-equilibrium distribution of quasiparticles[8], the photoelectric effect[9], microwave-assisted supercurrents[10], and the temperature for the onset of superconductivity[11,12]. The appearance of coherent excited states and the depairing effect of microwave radiation on dirty superconductors was very recently theoretically considered in ref. 13.

However, what remains virtually unexplored is how microwave radiation alters the superconducting proximity effect, which is the existence of superconducting correlations in an otherwise non-superconducting material when placed in contact with a superconductor, made possible due to electron tunneling between the layers. A concrete manifestation is the strong modification of the density of states, in both normal and magnetic metals proximity-coupled to a superconductor. The reason for why this is of importance is that proximity structures play a key part in creating non-conventional types of coherent electron pairing that are not present in ordinary superconductors. This includes both spin-polarized triplet superconductivity[14] and odd-frequency superconducting order[15], which recently have been experimentally demonstrated to provide diametrically opposite Meissner response[16] and low-energy spectral properties[17,18] compared to Bardeen-Cooper-Schrieffer theory[19]. From another perspective, the opportunity to manipulate low-energy excitations in superconducting proximity structures has clear practical implications for cryogenic technology since it controls the availability of spin- and charge-carriers. In fact, quasiparticles in superconductors can become nearly chargeless spin-1/2 carriers, leading to effects such as[20–24] strongly enhanced spin lifetimes and spin relaxation lengths when compared to injection of spin-polarized currents into normal metals, especially when using Zeeman split superconductors (a thin superconducting film in the presence of an in-plane magnetic field)[25]. This, in turn, allows one to envision various types of devices such as highly sensitive magneto- and thermometers as well as superconducting magnetoresistive elements.

Department of Physics, NTNU, Norwegian University of Science and Technology, N-7491 Trondheim, Norway. Correspondence and requests for materials should be addressed to J.L. (email: jacob.linder@ntnu.no)

In this work, we show that shining light on superconducting hybrid structures offers a way to control the proximity effect in both normal metals and magnetic materials. We discover that an oscillating electric field $\mathcal{E}(t)$ applied transversely to the junction induces a series of resonances in the density of states, and that it can be used to control the size of the minigap $E_{mg}$ in both bilayer superconductor/normal-metal (SN) and Josephson (SNS) junctions. The light interaction even inverts the minigap, generating a peak of quasiparticle accumulation at $E=0$ when the frequency of the light is tuned to $\omega = E_{mg}$. These findings give interesting prospects for transistor-like functionality via light-superconductor interactions since the density of states controls the availability of charge- and spin-carriers. Providing both analytical and numerical results, including the case of a magnetic exchange-field being present in the metal or in the superconductor, we show how the interaction between light and Cooper pairs controls the low-energy density of states, offering a new way to manipulate superconducting correlations. This may open a new pathway to active control of quantum coherent phenomena in superconducting proximity structures.

## Theory

We use the time-dependent quasiclassical Keldysh-Usadel theory[26–29] to describe the superconductivity of these systems in the diffusive limit. We begin with the SN bilayer, in which case superconducting correlations leak into the normal metal via the proximity effect. The electric field $\mathcal{E}(t) = \omega A_0 \sin(\omega t) = -\partial A/\partial t$ is accounted for by the gauge field $A = A_0 \cos(\omega t)$. The Usadel equation in N then reads:

$$D\partial_x(\hat{g}\partial_x\hat{g}) + i\left[E\hat{\rho}_3 + i\alpha\hat{\rho}_3(\hat{g}_+ + \hat{g}_-)\hat{\rho}_3, \hat{g}\right] = 0. \quad (1)$$

Here, $D$ is the diffusion coefficient, $\hat{g} = \hat{g}(x, E)$ is the quasiclassical time-averaged Green function, $E$ is the quasiparticle energy, $\alpha = DA_0^2/4$ is a measure of the strength of the interaction with light, $\omega$ is the driving frequency, $\hat{\rho}_3 = \text{diag}(+1, +1, -1, -1)$, while $\hat{g}_\pm \equiv \hat{g}(x, E \pm \omega/2)$. The derivation of this equation is shown in the Methods section and is valid when $\alpha \ll \omega$. We assume that the field is screened in the S region, which is taken to have a size and thickness far exceeding the superconducting coherence length $\xi$ and penetration depth $\lambda$, allowing us to use the bulk superconducting Green function $\hat{g}_{BCS}$ there. Practically, our proposed setup could be realized by depositing a thick superconductor to partially cover a thin normal metal layer, such that the microwave field penetrates the normal layer where it is not covered by a superconductor whereas it is shielded in the superconductor (see the inset of e.g. Fig. 1). Such a lateral geometry should be well described by an effective 1D model, as done in ref. 30. The thickness of the N layer should be much smaller than the skin depth and penetration depth $\lambda$, which is experimentally feasible (typical values for the skin depth of a normal metal such as Cu is of order µm at microwave frequencies, whereas $\lambda_{Nb} \sim 50$ nm and $\lambda_{Al} \sim 20$ nm). From Eq. (1), we derive the following Ricatti-parametrized[31,32] Usadel equation:

$$D[\partial_x^2\gamma + 2(\partial_x\gamma)\widetilde{\mathcal{N}}\tilde{\gamma}(\partial_x\gamma)] + 2i(E + i\delta)\gamma + i(\mathbf{h}\cdot\boldsymbol{\sigma})\gamma - i\gamma(\mathbf{h}\cdot\boldsymbol{\sigma}^*) - \alpha G\gamma - \alpha\gamma\widetilde{G} + \alpha(F + \gamma\widetilde{F}\gamma) = 0. \quad (2)$$

The Green function $\hat{g}$ can then be calculated from the $2\times 2$ matrix $\gamma$ in spin space, the normalization matrix $\mathcal{N} \equiv (1 - \gamma\tilde{\gamma})^{-1}$, and their tilde-conjugates defined by $\tilde{f}(x, E) \equiv f^*(x, -E)$. An equivalent equation for $\tilde{\gamma}$ can be found by tilde-conjugation of Eq. (2). In Eq. (2), we have also incorporated the possibility of a magnetic exchange field $h = |\mathbf{h}|$ which allows us to later consider the case of a ferromagnetic metal. The other quantities in the equation are the inelastic scattering rate $\delta$, and the short-hand notations

$$G \equiv \sum_\pm \mathcal{N}_\pm(1 + \gamma_\pm\tilde{\gamma}_\pm), \quad F \equiv -2\sum_\pm \mathcal{N}_\pm\gamma_\pm, \quad (3)$$

where $\gamma_\pm \equiv \gamma(x, E \pm \omega/2)$. From these equations, physical quantities of interest may be computed, such as the proximity-modified density of states

$$N/N_0 = \text{Re}\{\text{Tr}(\mathcal{N})\} - 1. \quad (4)$$

The Usadel equation is supplemented by the Kupriyanov-Lukichev boundary conditions[33], which are valid at low-transparency tunneling interfaces.

We now have at hand a coupled set of non-linear partial differential equations which are non-local in energy space. A numerical solution can be obtained via iteration. After discretizing the energy space, the equations are initially solved for $\alpha = 0$. The procedure is then repeated with $\alpha \neq 0$ until self-consistency is achieved, using the solutions $\gamma$ and $\tilde{\gamma}$ from the previous iteration to approximate $G$ and $F$. In this way, we are able to compute the quasiclassical Green function in the presence of microwave radiation, $\hat{g}(\alpha \neq 0)$, and access the density of states $N/N_0$ in the proximate metal.

## Results and Discussion

The light-interaction with the proximity-induced condensate has a strong effect on the spectral properties of the quasiparticles. We show this in what follows, considering an SN bilayer in Fig. 1, an SNS junction in Fig. 2, and an SF bilayer in Fig. 3. In each case, we have provided results for different system parameters in order to demonstrate the robustness of the microwave radiation influence.

Starting with the SN bilayer, it is seen that by tuning the microwave frequency $\omega$, the density of states takes on qualitatively different characteristics. At $\omega/\Delta_0 = 0.4$, there is a strong quasiparticle accumulation at $E = 0$, diametrically opposite to the hallmark minigap that usually is present in SN bilayers. Increasing $\omega$ gradually to $\omega/\Delta_0 = 1.0$ causes the density of states to revert to a minigap structure, albeit with a much reduced magnitude. We will later in this manuscript describe the precise condition leading to the appearance of the quasiparticle accumulation peak and its physical origin, providing also analytical results which supports the underlying explanation. In the plots, we have set $\alpha/\Delta_0 = 0.1$, which gives a maximum ratio of $\alpha/\omega = 0.25$, so that $\alpha$ is always considerably smaller than $\omega$. The criterion $\alpha \ll \omega$ is, however, more strictly satisfied at the higher frequency range considered in the figures.

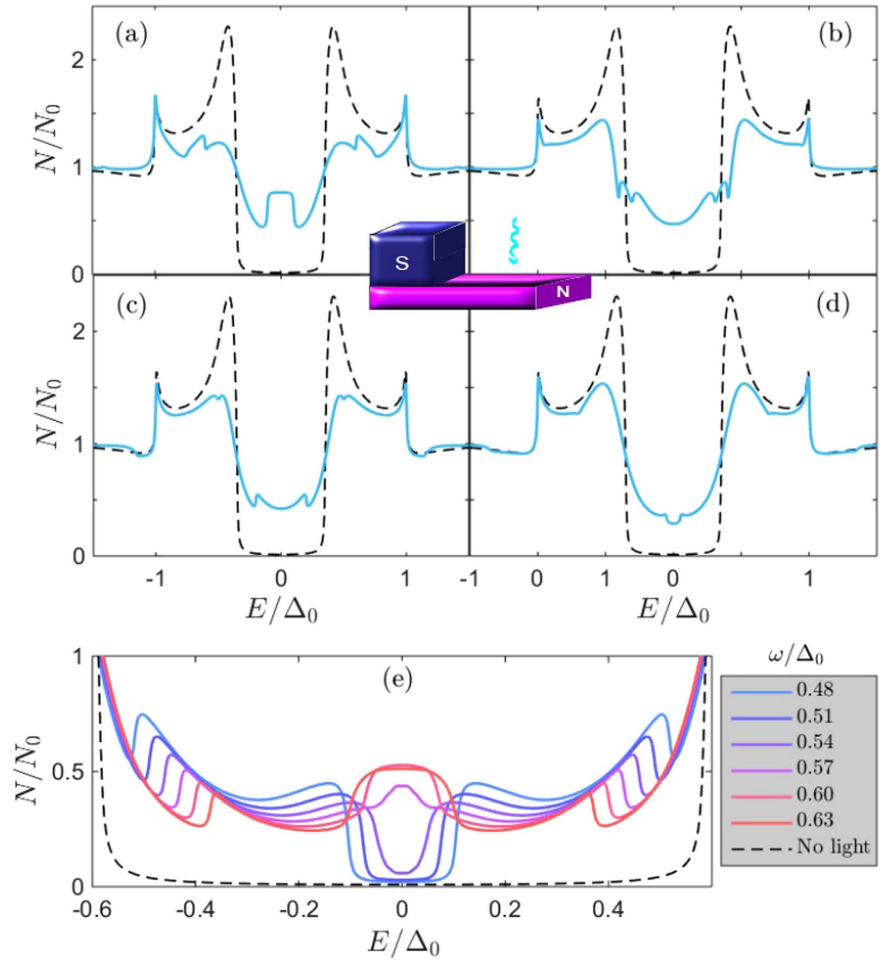

**Figure 1.** (**a**–**d**) Proximity-induced density of states at the vacuum edge ($x = L$) of an SN bilayer with length $L/\xi = 0.5$ of the N region, where $\xi$ is the superconducting coherence length. We set the barrier strength $\zeta = 3$ and microwave field amplitude $\alpha/\Delta_0 = 0.1$ and (**a**) $\omega/\Delta_0 = 0.4$, (**b**) $\omega/\Delta_0 = 0.6$, (**c**) $\omega/\Delta_0 = 0.8$, (**d**) $\omega/\Delta_0 = 1.0$. (**e**) Zoom-in near $E = 0$ illustrating the transition from minigap to quasiparticle accumulation peak as $\omega$ is tuned to $E_{mg}$. We set $L/\xi = 0.33$, yielding $E_{mg} \simeq 0.56\Delta_0$. The black dashed line corresponds to the absence of light, $A = 0$.

The minigap itself is monotonically tuned with $\omega$, as shown in Fig. 2 for the SNS case. At zero phase difference $\phi$, the minigap is gradually reduced as $\omega$ increases, demonstrating that the driving frequency can be used to tailor the minigap size. At a finite phase difference, the light-interaction again inverts the minigap for certain frequencies, and generates a peak of quasiparticle accumulation at $E = 0$, similarly to the bilayer case [see Fig. 1(e)]. This can be seen in Fig. 2(c) for $\phi/\pi = 0.5$. Finally, we show results for when an exchange field is present, i.e. a magnetic metal $h \ne 0$, in Fig. 3, in which case the microwave field also alters the modulation of the density of states. To facilitate comparison with experiments, we note that for a typical diffusion constant of e.g. $D = 7 \times 10^{-3}$ m²/s in Cu[34], the requirement $\omega \gg De^2 A_0^2/4\hbar^2$ (having reinstated $e$ and $\hbar$) corresponds to $\omega \gg 0.3$ GHz for a modest electric field magnitude of 0.1 V/m, which is feasible. Moreover, for a superconducting gap $\Delta_0 = 0.5$ meV, the parameter choice $\hbar\omega/\Delta_0 = 0.4$ corresponds to a frequency $\omega \simeq 300$ GHz.

Besides the control and inversion of the minigap, another particularly noteworthy feature that all the above-mentioned structures have in common is that the low-energy density of states features a series of spectral features resembling weak resonances, which vanish as soon as the microwave field is turned off ($\alpha = 0$). To gain insight into the physical origin of these features seen in the density of states, we provide an analytical solution which is permissible in the ferromagnetic case, but which also seems to account for the nature of the light interaction with the superconducting condensate in the normal case ($h = 0$). In the weak proximity effect regime, the linearized equation governing the behavior of the spinless $f_s$ and spin-polarized $f_t$ Cooper pairs reads

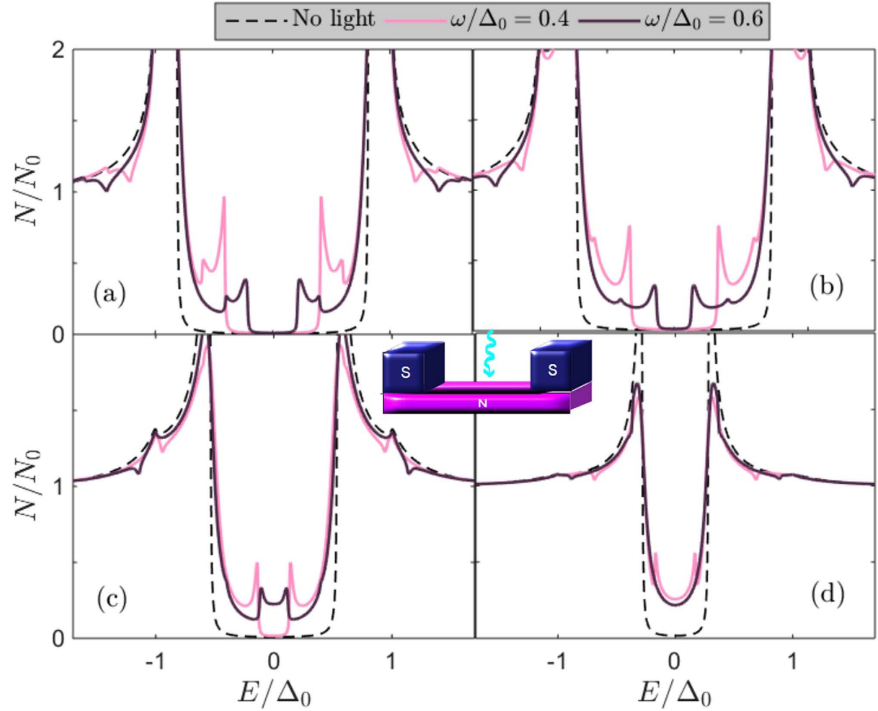

**Figure 2.** Proximity-induced density of states in the middle ($x = L/2$) of an SNS Josephson junction with $L/\xi = 0.33$, barrier strength $\zeta = 3$, microwave field amplitude $\alpha/\Delta_0 = 0.1$, and (**a**) $\phi/\pi = 0.0$, (**b**) $\phi/\pi = 0.25$, (**c**) $\phi/\pi = 0.5$, (**d**) $\phi/\pi = 0.75$.

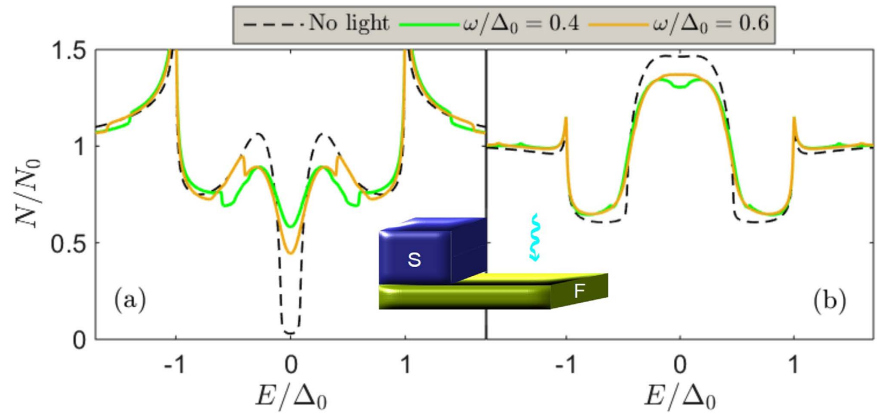

**Figure 3.** Proximity-induced density of states at the vacuum edge ($x = L$) of an SF bilayer with $L/\xi = 0.23$, barrier strength $\zeta = 3$, microwave field amplitude $\alpha/\Delta_0 = 0.1$, and (**a**) $h/\Delta_0 = 2$ and (**b**) $h/\Delta_0 = 4$.

$$D\partial_x^2 f_\pm(E) + 2i(E + 2i\alpha \pm h) f_\pm(E) - 2\alpha [f_\pm(E + \omega) + f_\pm(E - \omega)] = 0 \quad (5)$$

with $f_\pm = f_t \pm f_s$. In the regime where $h \gg \{E, \Delta_0\}$, as is usually the case for ferromagnets, one can solve the above equation via Fourier-transformation. Introducing $\mathcal{F}_\pm(t) = \int dE e^{iEt} f_\pm(E)$, one obtains

$$D\partial_x^2 \mathcal{F}_\pm(t) + 2i(2i\alpha \pm h) \mathcal{F}_\pm(t) - 4\alpha \cos(\omega t) \mathcal{F}_\pm(t) = 0.$$

The solution is $\mathcal{F}_\pm(t) = A_\pm(t) e^{ik_\pm x} + B_\pm(t) e^{-ik_\pm x}$, where

$$k_\pm = \sqrt{[2i(2i\alpha \pm h) - 4\alpha \cos(\omega t)] D^{-1}}, \quad (6)$$

while the coefficients $\{A_\pm, B_\pm\}$ are determined via the boundary conditions. For an SF bilayer, the boundary conditions read $\partial_x f_\pm = \pm f_{\text{BCS}}/\zeta L$ at the superconducting interface ($x = 0$), and $\partial_x f_\pm = 0$ at the vacuum border ($x = L$), where $\zeta = R_B/R$ is the ratio between the interface barrier resistance and bulk resistance and $f_{BCS}(E) = \sinh\{\text{atanh}[1/(E + i\delta)]\}$.

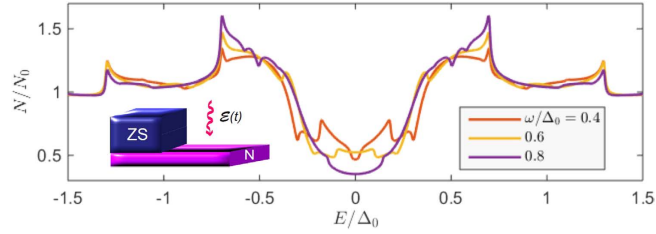

**Figure 4. Proximity-induced density of states at the vacuum edge ($x = L$) of a Zeeman-split superconductor/normal-metal bilayer.** We set $\zeta = 3$, $\alpha/\Delta_0 = 0.1$, $L/\xi = 0.5$, and $h_s/\Delta_0 = 0.3$, and considered several frequencies of the microwave radiation.

Introducing the auxiliary quantity $D(t) = \int dE\, e^{iEt} f_{BCS}(E)$, a straight-forward calculation leads to

$$A_{\pm}(t) = \frac{\mp D(t)}{\zeta L i k_{\pm}(1 - e^{2ik_{\pm}L})}, \quad B_{\pm}(t) = A_{\pm}(t) e^{2ik_{\pm}L}. \tag{7}$$

Inserting this into our expression for $\mathcal{F}_{\pm}(t)$ and performing an inverse Fourier-transformation, we end up with the final expression for $f_{\pm}(E)$:

$$f_{\pm}(E) = \mp \iint dt\, dE'\, e^{i(E'-E)t} f_{BCS}(E') p_{\pm}(t). \tag{8}$$

where we introduced

$$p_{\pm}(t) = \cos[k_{\pm}(x - L)][\zeta L k_{\pm} \sin(k_{\pm} L)]^{-1} \tag{9}$$

and $k_{\pm} = k_{\pm}(t)$. We note that $p_{\pm}(t)$ is a periodic function in $t$, while $f_{BCS} \to 0$ when $E \to \pm\infty$. In the absence of microwave radiation ($\alpha = 0$), $k_{\pm}$ becomes independent of $t$, and the above simplifies to the usual result $f_{\pm}(E) = \mp p_{\pm} \iint dt\, dE'\, e^{i(E'-E)t} f_{BCS}(E') = \mp p_{\pm} f_{BCS}(E)$. To solve the integral Eq. (8) in the general case, we make use of the periodicity of $p_{\pm}(t)$. The period is $T = 2\pi/\omega$, so we can write the Fourier series $p_{\pm}(t) = \sum_n p_{n,\pm} e^{in\omega t}$, where $p_{n,\pm} = \frac{1}{T} \int_{-T/2}^{+T/2} dt\, p_{\pm} e^{-in\omega t}$. Performing the integral over $t$ in Eq. (8) then leads to a sum over $\delta$-functions, and one obtains:

$$f_{\pm}(E) = \mp \sum_{n=-\infty}^{n=+\infty} p_{n,\pm} f_{BCS}(E - n\omega). \tag{10}$$

Numerically, we find that it is usually sufficient with ~15 Fourier-coefficients $p_{n,\pm}$ to obtain a perfect representation of $p_{\pm}(t)$. Using the same procedure as above, one can also find an expression for the anomalous Green function in a Josephson geometry consisting of a superconductor/ferromagnet/superconductor trilayer. The only difference is the expression for $p_{\pm}(t)$, which takes the form

$$p_{\pm}(t) = \{[\cos(k_{\pm} x) + e^{i\phi} \cos[k_{\pm}(x - L)]][\zeta L k_{\pm} \sin(k_{\pm} L)]^{-1}, \tag{11}$$

where $\phi$ is the phase difference between the superconductors.

From the analytical expression, it is clear that resonances should be expected whenever $E = \Delta_0 \pm n\omega$, $n = 0, 1, 2, \ldots$ since $f_{BCS}(E = \Delta_0)$ formally diverges, although this divergence is in practice diminished due to inelastic scattering. The weight of these resonances, i.e. the magnitude of their spectral peak, is in turn governed by the Fourier series coefficients $p_n$ which depends on the other system parameters. We note that, very recently, similar features were reported for a narrow and thin dirty superconducting strip subject to microwave radiation in ref. 13. In the present proximity-system, there is an additional minigap $E_{mg}$ in the system, and one might expect to have similar resonances at $E = E_{mg} \pm n\omega$. The density of states plots in Fig. 2 [see for instance (a) for $\omega/\Delta_0 = 0.4$] are consistent with this statement, demonstrating how additional spectral features, which are not present in the absence of light, occur at such excitation energies. It actually turns out that these resonances are the physical origin behind the transition from the minigap to the quasiparticle accumulation peak at $E = 0$. To be exact, the transition from fully gapped DOS to a strong zero-energy peak occurs precisely when $\omega = E_{mg}$. We show an example of this behavior at the bottom of Fig. 1. It is intriguing that the light-interaction actually induces a second, inner minigap which upon closing generates this feature, whereas the outer minigap $E_{mg}$ remains [see e.g. Fig. 2(a) showing a particularly clear example of the inner and outer minigaps].

The fact that the microwave radiation induces a series of weak resonances shifted with $\pm n\omega$ from the conventional spectral peaks ($E = \Delta$ and $E = E_{mg}$ in the normal metal case) has interesting consequences when a finite magnetic field splits the density of states in the superconductor[35], since the exchange field in the superconductor $h_S$ itself produces a similar shift in the spectral peaks from $\Delta_0$ to $\Delta_0 \pm h_S$. We show the corresponding proximity-induced density of states in Fig. 4, where the combined influence of the exchange field and the light interaction produce a very rich subgap structure in the density of states. Since the superconductor in this particular case, unlike the previous systems considered in this work, has to be sufficiently thin to permit the homogeneous penetration of a magnetic field, the microwave field is not completely shielded by the superconductor and we thus here assumed that $\mathcal{E}(t)$ is applied only to the non-superconducting part.

The most remarkable feature is nevertheless the influence of the microwave field on the minigap in the SN case, controlling its magnitude and even transforming it into a quasiparticle accumulation peak at $E = 0$. These results may represent the first step toward a different way to control the superconducting proximity effect, and thus the available spin- and charge-carriers, in normal and magnetic metals, by using microwave radiation. One advantage of this is the fact that the control is *in situ* and that the length of the system (setting the Thouless energy scale), which normally changes the minigap, does not have to be altered, which would inevitably require fabrication of multiple samples. The zero-energy peak induced by the light-interaction resembles the type of spectral feature that is characteristically seeen in the density of states of conventional SF structures due to odd-frequency superconductivity[36–38], but in this case it occurs without any such pairing at all. It could also be of interest to examine the consequences of the predictions made herein with regard to conductivity experiments[39] and non-equilibrium Josephson contacts[40].

## Concluding remarks

Building on these results, an interesting future direction to explore would be the influence of light on supercurrents and the critical temperature in magnetic proximity systems, to see if the microwave radiation may be used to manipulate these quantities as well, which we intend to explore in a future work. The interaction between light and Cooper pairs could in this way open a different route to active control of quantum coherent phenomena in superconducting proximity structures.

## Methods

### Derivation of the Usadel equation incorporating microwave radiation.
The time-dependent Usadel equation may be written as

$$D\overline{\nabla} \circ (\hat{g} \circ \overline{\nabla} \circ \hat{g}) = -i[E\hat{\rho}_3 \overset{\circ}{,} \hat{g}], \tag{12}$$

where we defined the gauge-covariant derivative

$$\overline{\nabla} \circ \hat{g} \equiv \nabla \hat{g} - ie[A\hat{\rho}_3 \overset{\circ}{,} \hat{g}], \tag{13}$$

the commutator $[a \overset{\circ}{,} b] \equiv a \circ b - b \circ a$, and the associated product

$$(a \circ b)(E, T) \equiv e^{i(\partial_{E_1}\partial_{T_2} - \partial_{E_2}\partial_{T_1})/2} a(E_1, T_1) \times b(E_2, T_2) \Big|_{E_1 = E_2 = E, T_1 = T_2 = T}. \tag{14}$$

Above, $e$ is the electron charge, $E$ is the quasiparticle energy, and $\boldsymbol{A}$ is the time-dependent vector potential which describes, in our case, an ac electric field $\boldsymbol{E} = -\partial \boldsymbol{A}/\partial t$. We note that a useful property of the ∘-product is that:

$$a(E, T) \circ e^{i\omega T} = e^{i\omega T} a(E - \omega/2, T), \quad e^{i\omega T} \circ a(E, T) = e^{i\omega T} a(E + \omega/2, T). \tag{15}$$

These relations are useful in the present context since we can write the gauge field as

$$A(T) = A_0(e^{i\omega T} + e^{-i\omega T})/2. \tag{16}$$

We set $|e| = 1$ in what follows for brevity of notation and also apply the electric field perpendicularly to the junction direction, so that

$$\nabla \cdot \boldsymbol{A} = \boldsymbol{A} \cdot \nabla = 0. \tag{17}$$

In this case, the left hand side of Eq. (12) becomes

$$D\nabla \cdot (\hat{g} \circ \nabla \hat{g}) - D[A\hat{\rho}_3 \overset{\circ}{,} \hat{g} \circ A\hat{\rho}_3 \circ \hat{g} - A\hat{\rho}_3]. \tag{18}$$

Since $\boldsymbol{A} = \boldsymbol{A}(T)$ is independent on $E$ we have

$$A\hat{\rho}_3 A\hat{\rho}_3 = A^2 \hat{1}. \tag{19}$$

Moreover, the Green function satisfies the normalization condition

$$\hat{g} \circ \hat{g} = \hat{1}. \tag{20}$$

This brings us to

$$D\nabla \cdot (\hat{g} \circ \nabla \hat{g}) - D[A\hat{\rho}_3 \circ \hat{g} \circ A\hat{\rho}_3 \overset{\circ}{,} \hat{g}]. \tag{21}$$

At this stage, we see that the contribution from the gauge field can be included as a self-energy

$$\hat{\Sigma}_A = iD\mathbf{A}\hat{\rho}_3 \circ \hat{g} \circ \mathbf{A}\hat{\rho}_3 \qquad (22)$$

in the Usadel equation, which in its complete form reads:

$$D\nabla \cdot (\hat{g} \circ \nabla \hat{g}) + i[E\hat{\rho}_3 + iD\mathbf{A}\hat{\rho}_3 \circ \hat{g} \circ \mathbf{A}\hat{\rho}_3, \hat{g}] = 0. \qquad (23)$$

The next step is the obtain the Fourier-transformed version of the above equation in energy-space. To accomplish this, we make use of similar approximations as in ref. 13. In the presence of a driving field $A(T)$, we take into account $A$ up to second order by deriving an equation for the harmonic Green function at zero frequency (see Appendix of ref. 13) which is essentially the time-averaged Green function. Higher order harmonic time-dependent terms in $\hat{g}$ are induced by $A$ and thus correspond to fourth order in $A$ and higher. This approximation is valid when

$$DA_0^2/4 \ll \omega. \qquad (24)$$

Computing the contribution from the self-energy term $\hat{\Sigma}_A$ in the Usadel equation gives

$$i[\hat{\Sigma}_A, \hat{g}] = -\frac{DA_0^2}{4}[(e^{i\omega T} + e^{-i\omega T})\hat{\rho}_3 \circ \hat{g} \circ (e^{i\omega T} + e^{-i\omega T})\hat{\rho}_3 \circ \hat{g}$$
$$- \hat{g} \circ (e^{i\omega T} + e^{-i\omega T})\hat{\rho}_3 \circ \hat{g} \circ (e^{i\omega T} + e^{-i\omega T})\hat{\rho}_3]. \qquad (25)$$

We now average Eq. (25) over a period $2\pi/\omega$, which means that all terms that go like $e^{\pm 2i\omega T}$ are removed since $\hat{g}$ is the time-averaged Green function. After laborious calculations, using for instance that

$$e^{i\partial_{E_1}\partial_{T_2}/2}\left[1 - \frac{i}{2}\partial_{E_2}\partial_{T_1} + \frac{1}{2}\left(\frac{i}{2}\right)^2(\partial_{E_2}\partial_{T_1})^2 - \ldots\right]e^{\pm i\omega T_1}\hat{\rho}_3\hat{g}(E_1 \pm \omega/2)e^{\mp i\omega T_2}$$
$$\times \hat{\rho}_3\hat{g}(E_2 \mp \omega/2)\big|_{E_1=E_2=E, T_1=T_2=T}.$$
$$= e^{i\partial_{E_1}\partial_{T_2}/2}\left[1 \mp \frac{i}{2}(i\omega)\partial_{E_2} + \frac{1}{2}\left(\frac{\omega}{2}\right)^2(\partial_{E_2})^2 - \ldots\right]e^{\pm i\omega T_1}$$
$$\times \hat{\rho}_3\hat{g}(E_1 \pm \omega/2)e^{\mp i\omega T_2}\hat{\rho}_3\hat{g}(E_2 \mp \omega/2)\big|_{E_1=E_2=E, T_1=T_2=T}.$$
$$= e^{i\partial_{E_1}\partial_{T_2}/2}e^{\pm i\omega T_1}e^{\mp i\omega T_2}\hat{\rho}_3\hat{g}(E_1 \pm \omega/2)\hat{\rho}_3\hat{g}(E_2)\big|_{E_1=E_2=E, T_1=T_2=T}.$$
$$= \hat{\rho}_3\hat{g}(E \pm \omega)\hat{\rho}_3\hat{g}(E) \qquad (26)$$

via Eq. (15), the remaining terms take the form

$$D\nabla \cdot (\hat{g}\nabla\hat{g}) + i\left[E\hat{\rho}_3 + i\alpha\hat{\rho}_3(\hat{g}_+ + \hat{g}_-)\hat{\rho}_3, \hat{g}\right] = 0, \qquad (27)$$

where the $\circ$-commutators are now replaced with regular matrix commutators, $\alpha \equiv DA_0^2/4$, $\hat{g} = \hat{g}(x, E)$ is the quasiclassical Green function, while

$$\hat{g}_\pm \equiv \hat{g}(x, E \pm \omega/2). \qquad (28)$$

**Derivation of the linearized Usadel equation (weak proximity effect).** Analytical progress can be made in the so-called weak proximity effect regime, where one assumes that the magnitude of the superconducting proximity effect is small in the sense that the anomalous Green function components $f$ satisfy $|f| \ll 1$. Physically, such a situation is realized either in the case of a low interface transparency between the superconducting and normal part or if the temperature is close to the critical temperature of the superconductor. This allows for a linearization of the Usadel equation in the anomalous Green functions in the following manner[3]. The total Green function matrix in Nambu-spin space may be written as the normal-state matrix $\hat{g}_0$ and a small deviation $\hat{f}$:

$$\hat{g} \simeq \hat{g}_0 + \hat{f}, \qquad (29)$$

where $\hat{g}_0 = \hat{\rho}_3$ and the anomalous Green function matrix can be written as

$$\hat{f} = \begin{pmatrix} \underline{0} & \underline{f} \\ -\underline{\tilde{f}} & \underline{0} \end{pmatrix}. \qquad (30)$$

The $2 \times 2$ matrix $\underline{f}$ in spin space describes the four types of anomalous Green functions that can be present in the system: one describing spin-singlet Cooper pairs ($f_s$) and three describing spin-triplet Cooper pairs ($f_{\uparrow\uparrow}, f_{\downarrow\downarrow}, f_t$). The $f_t$ component corresponds to the $S = 1$, $S_z = 0$ component of the triplets with spin-symmetry $\uparrow\downarrow + \downarrow\uparrow$ and the $\sim$ operation is defined in the main text. For the systems considered in our work, with homogene-ous exchange fields, we find that $f_{\sigma\sigma} = 0$ whereas $f_s$ and $f_t$ can be non-zero. Inserting Eq. (29) into Eq. (2) in the main manuscript produces the linearized equation

$$D\partial_x^2 f_\pm(E) + 2i(E + 2i\alpha \pm h)f_\pm(E) - 2\alpha[f_\pm(E+\omega) + f_\pm(E-\omega)] = 0 \quad (31)$$

with $f_\pm = f_t \pm f_s$. This governs the behavior of the spinless $f_s$ and spin-polarized $f_t$ Cooper pairs induced in the normal metal.

## References

1. J. Linder & J. W. A. Robinson. Superconducting spintronics. *Nat. Phys.* **11,** 307 (2015).
2. J. W. A. Robinson & M. G. Blamire. The interface between superconductivity and magnetism: understanding and device prospects. *J. Phys.: Cond. Mat.* **26,** 453201 (2014).
3. M. Eschrig. Spin-polarized supercurrents for spintronics: a review of current progress. *Rep. Prog. Phys.* **78,** 10 (2015).
4. I. O. Kulik. Nonlinear High-frequency Properties of Thin Superconducting Films. *Sov. Phys. JETP* **30,** 329 (1970).
5. A. Gurevich. Reduction of Dissipative Nonlinear Conductivity of Superconductors by Static and Microwave Magnetic Fields. *Phys. Rev. Lett.* **113,** 087001 (2014).
6. F. S. Bergeret, P. Virtanen, T. T. Heikkilä & J. C. Cuevas. Theory of Microwave-Assisted Supercurrent in Quantum Point Contacts. *Phys. Rev. Lett.* **105,** 117001 (2010).
7. F. Kos, S. E. Nigg & L. I. Glazman. Frequency-dependent admittance of a short superconducting weak link. *Phys. Rev. B* **87,** 174521 (2013).
8. P. Virtanen, T. T. Heikkilä & F. S. Bergeret. Stimulated quasiparticles in spin-split superconductors. *Phys. Rev. B* **93,** 014512 (2016).
9. M. S. Kalenkov & A. D. Zaikin. Diffusive superconductors beyond the Usadel approximation: Electron-hole asymmetry and large photoelectric effect. *Phys. Rev. B* **92,** 014507 (2015).
10. P. Virtanen *et al.* Theory of Microwave-Assisted Supercurrent in Diffusive SNS Junctions. *Phys. Rev. Lett.* **104,** 247003 (2010).
11. J. E. Mooij. In *Nonequilibrium Superconductivity, Phonons and Kapitza Boundaries*, edited by K. E. Gray (Plenum, New York, 1981).
12. V. M. Dmitriev, V. N. Gubankov & F. Y. Nad. In *Nonequilibrium Superconductivity*, edited by D. N. Langenberg & A. I. Larkin (North-Holland, Amsterdam 1986).
13. A. V. Semenov, I. A. Devyatov, P. J. de Visser & T. M. Klapwijk. Coherent Excited States in Superconductors due to a Microwave Field. *Phys. Rev. Lett.* **117,** 047002 (2016).
14. A. P. Mackenzie & Y. Maeno. The superconductivity of $Sr_2RuO_4$ and the physics of spin-triplet pairing. *Rev. Mod. Phys.* **75,** 657 (2003).
15. V. L. Berizinskii. New model of the anisotropic phase of superfluid $He^3$. *JETP Lett.* **20,** 287 (1974).
16. A. Di Bernardo *et al.* Intrinsic Paramagnetic Meissner Effect Due to *s*-Wave Odd-Frequency Superconductivity. *Phys. Rev. X* **5,** 041021 (2015).
17. A. Di Bernardo *et al.* Signature of magnetic-dependent gapless odd frequency states at superconductor/ferromagnet interfaces. *Nat. Commun.* **6,** 8053 (2015).
18. Y. Kalcheim, O. Millo, A. Di Bernardo, A. Pal & J. W. A. Robinson. Inverse proximity effect at superconductor-ferromagnet interfaces: Evidence for induced triplet pairing in the superconductor. *Phys. Rev. B* **92,** 060501(R) (2015).
19. J. Bardeen, L. N. Cooper & J. R. Schrieffer. Theory of Superconductivity. *Phys. Rev.* **108,** 1175 (1957).
20. T. Yamashita, S. Takahashi, H. Imamura & S. Maekawa. Spin transport and relaxation in superconductors. *Phys. Rev. B* **65,** 172509 (2002).
21. H. Yang, S.-H. Yang, S. Takahashi, S. Maekawa & S. S. P. Parkin. Extremely long quasiparticle spin lifetimes in superconducting aluminium using MgO tunnel spin injectors. *Nature Mater.* **9,** 586 (2010).
22. C. H. L. Quay, D. Chevallier, C. Bena & M. Aprili. Spin imbalance and spin-charge separation in a mesoscopic superconductor. *Nature Phys.* **9,** 84 (2013).
23. F. Hübler, M. J. Wolf, D. Beckmann & H. v. Löhneysen. Long-range spin-polarized quasiparticle transport in mesoscopic Al superconductors with a Zeeman splitting. *Phys. Rev. Lett.* **109,** 207001 (2012).
24. T. Wakamura, N. Hasegawa, K. Ohnishi, Y. Niimi & Y. Otani. Spin injection into a superconductor with strong spin-orbit coupling. *Phys. Rev. Lett.* **112,** 036602 (2014).
25. D. Beckmann. Spin manipulation in nanoscale superconductors. *J. Phys. Cond. Mat.* **28,** 163001 (2016).
26. J. Rammer & H. Smith. Quantum field-theoretical methods in transport theory of metals. *Rev. Mod. Phys.* **58,** 323 (1986).
27. W. Belzig, F. K. Wilhelm, C. Bruder, G. Schön & A. D. Zaikin. Quasiclassical Green's function approach to mesoscopic superconductivity. *Superlattices Microstruct.* **25,** 1251 (1999).
28. V. Chandrasekhar. *An introduction to the quasiclassical theory of superconductivity for diffusive proximity-coupled systems*, Chapter published in, "The Physics of Superconductors", Vol II, edited by Bennemann & Ketterson, Springer-Verlag (2004).
29. K. Usadel. Generalized Diffusion Equation for Superconducting Alloys. *Phys. Rev. Lett.* **25,** 507 (1970).
30. H. le Sueur *et al.* Phase Controlled Superconducting Proximity Effect Probed by Tunneling Spectroscopy. *Phys. Rev. Lett.* **100,** 197002 (2008).
31. N. Schopohl & K. Maki. Quasiparticle spectrum around a vortex line in a d-wave superconductor. *Phys. Rev. B* **52,** 490 (1995).
32. N. Schopohl. Transformation of the Eilenberger Equations of Superconductivity to a Scalar Riccati Equation. arXiv:cond-mat/9804064.
33. M. Y. Kupriyanov & V. F. Lukichev. Influence of boundary transparency on the critical current of "dirty" SS'S structures. *Sov. Phys. JETP* **67,** 1163 (1988).
34. S. Gueron *et al.* Superconducting Proximity Effect Probed on a Mesoscopic Length Scale. *Phys. Rev. Lett.* **77,** 3025 (1996).
35. R. Meservey, P. M. Tedrow & P. Fulde. Magnetic Field Splitting of the Quasiparticle States in Superconducting Aluminum Films. *Phys. Rev. Lett.* **25,** 1270 (1970).
36. T. Kontos, M. Aprili, J. Lesueur & X. Grison. Inhomogeneous Superconductivity Induced in a Ferromagnet by Proximity Effect. *Phys. Rev. Lett.* **86,** 304 (2001).
37. T. Yokoyama, Y. Tanaka & A. A. Golubov. Manifestation of the odd-frequency spin-triplet pairing state in diffusive ferromagnet/superconductor junctions. *Phys. Rev. B* **75,** 134510 (2007).
38. J. Linder, A. Sudbø, T. Yokoyama, R. Grein & M. Eschrig. Signature of odd-frequency pairing correlations induced by a magnetic interface. *Phys. Rev. B* **81,** 214504 (2010).
39. M. S. Pambianchi, S. N. Mao & S. M. Anlage. Microwave surface impedance of proximity-coupled Nb/Al bilayer films. *Phys. Rev. B* **52,** 4477 (1995).
40. J. J. A. Baselmans, A. F. Morpurgo, B. J. van Wees & T. M. Klapwijk. Reversing the direction of the supercurrent in a controllable Josephson junction. *Nature* **397,** 43 (1999).

### Acknowledgements
Funding via the "Outstanding Academic Fellows" programme at NTNU, the COST Action MP-1201, the NT-Faculty, and the Research Council of Norway Grant numbers 216700 and 240806, is gratefully acknowledged.

### Author Contributions
J.L. did the majority of the analytical and numerical calculations with support from M.A. and J.A.O. All authors contributed to the discussion of the results and the writing of the manuscript.

### Additional Information
**Competing financial interests:** The authors declare no competing financial interests.

**How to cite this article**: Linder, J. *et al.* Microwave control of the superconducting proximity effect and minigap in magnetic and normal metals. *Sci. Rep.* **6**, 38739; doi: 10.1038/srep38739 (2016).

**Publisher's note:** Springer Nature remains neutral with regard to jurisdictional claims in published maps and institutional affiliations.